\documentclass[namedreferences]{kluwer}

\usepackage{epsfig}

\begin{document}

\begin{article}

\begin{opening}

\title{Extrapolation of the solar magnetic field within the potential-field
approximation from full-disk magnetograms}


\author{G. V. \surname{Rudenko}\email{rud@iszf.irk.ru}}
\institute{Institute of Solar-Terrestrial Physics, Irkutsk, Russia}


\runningtitle{Extrapolation of the solar magnetic field from full-disk magnetograms}

\runningauthor{G.V. RUDENKO}

\begin{ao}
  Institute of Solar-Terrestrial Physics,
  P.O. Box 4026, Irkutsk, 664033, Russia \\
  e-mail: rud@iszf.irk.ru \\
  Fax: +7 (3952) 46 25 57
\end{ao}


\begin{abstract}
This paper is concerned with the Laplace boundary-value problem with the
directional derivative, corresponding to the specific nature of measurements
of the longitudinal component of the photospheric magnetic field. Boundary
conditions are specified by a distribution on the sphere of projection of
the magnetic field vector unto a given direction. It is shown that the
solution of this problem exists in the form of a spherical harmonic
expansion, and uniqueness of this solution is proved. A conceptual sketch of
numerical determination of the harmonic series coefficients is given. The
field of application of the method is analyzed having regard to the
peculiarities of actual data. Finally, we present differences in results
derived from extrapolating the magnetic field from a synoptic map and a
full-disk magnetogram.
\end{abstract}

\end{opening}

\section{Introduction}

Until the present time the practical potential-field extrapolation of solar
magnetic fields has been based on several thoroughly studied Laplace
boundary-value problems when applied to two cases of geometry: spherics, and
planar geometry. The spherics, as a rule, is used in investigating
large-scale long-lived global fields. The planar geometry is exploited to
study magnetic active regions occupying small areas on the solar surface.

When employing the spherics, two approaches are distinguished to specifying
boundary conditions on the solar surface. One is that it is specified the
synoptic distribution of a so-called component $B_l$ (a projection of the
magnetic field measured along the line-of-sight direction is assigned to
each point of the surface at the time when a given point passes through the
central meridian (\cite{Altschuler} and \cite{Hoeksema}). In the other
approach, a normal component $B_r$ (\cite{Wang and Sheeley}) is specified,
which is constructed from $B_l$ assuming that the full vector of the
magnetic field is radial. The latter case corresponds to the setting of a
classical Neumann boundary-value problem, and the former case refers to a
particular variant of the boundary-value problem with the directional
derivative (the direction of the derivative with respect to a normal varies
according to the heliographic latitude). Data of daily magnetograms do not
satisfy these conditions. They are only approximately satisfied by points
lying in the neighborhood of the central meridian. Synoptic maps when
generated from series of daily magnetograms accumulated during one solar
rotation provide input information corresponding to the above-mentioned
boundary conditions.

Using the planar geometry, unlike the spherics, provides insight into the
three-dimensional spatial structure at the time of measurement thus enabling
one to track the temporal evolution of magnetic active regions. This
approach has two important limitations. In the first place, the region under
study must be sufficiently small; secondly, it must lie in the middle of the
solar disk where the normality approximation of the measured component of
field is valid. Boundary conditions in this case correspond to the Neumann
problem. A variant of the problem with a constant slope of the derivative
(for noncentral regions) can also be used but at a still greater sacrifice
in the size of the region if variations in the line-of-sight slope to a
normal inside the region are to be neglected.

This paper offers the solution of a Laplace boundary-value problem (in the
spherics) with boundary conditions corresponding to measurements of the
longitudinal magnetic components throughout the disk (data from daily
magnetograms). Such a setting of the problem conforms to a Laplace problem
with a variable directional derivative (by choosing a rigorously specified
projection direction corresponding to the direction of the line of sight
existing at a given time of measurement, the direction, with respect to
which a derivative of the scalar magnetic potential is taken, changes
relative to a normal on the boundary spherical surface). This let to use
completely longitudinal magnetogram data which are the most sensitive and
generally applied. We demonstrate the existence of a strictly definite
harmonic expansion of the potential field in terms of spherical
eigenfunctions, corresponding to the above boundary conditions. The
construction of such a solution is based on the combined use of Galerkin's
projection method and Trefftz' method (the solutions are sought in a class
of solutions of the Laplace equation). The proof of the existence and
uniqueness of the solution is constructed. We present and analyze results of
numerical tests simulating the extrapolation of actual magnetograms. The
limitations of the approach used in this study are discussed, which are
associated with the resolution of the magnetograms employed and with the
uncertainty of specifying the magnetogram of the averted side of the Sun
(''back magnetogram''). A comparison is made of results derived from
extrapolating from a synoptic magnetogram and from a full-disk magnetogram.

\section{Statement of the boundary-value problem}

Assume that a force-free currentless approximation holds either throughout
the space above the photospheric surface $S_p$, or between the photosphere
and some spherical surface $S_s$ (source surface). Consider a boundary-value
problem
\begin{equation}
\bf{B}\left( \bf{r}\right) =-\bf{\nabla }\Psi \left( \bf{r}%
\right)  \label{a1}
\end{equation}
\begin{equation}
\Delta \Psi \left( \bf{r}\right) =0,\bf{r\in }\Omega \bf{%
,\partial }\Omega \bf{=}S  \label{a2}
\end{equation}

\begin{equation}
-\bf{d}\cdot \bf{\nabla }\Psi \left( \bf{r}\right) =B_d(\theta
,\varphi )\ \ \ \ \mathit{for}\ \ \bf{r\in }S_p  \label{a3}
\end{equation}

\begin{equation}
\left\{
\begin{array}{c}
\ a)\;\Psi \left( \bf{r}\right) =0\ \ \ \ \mathit{for}\ \ \bf{r\in }%
S_s,S=S_p\cup S_s, \\
b)\;\Psi \left( \bf{r}\right) \propto O(1/r),\ r\rightarrow \infty ,\
S=S_p\qquad
\end{array}
\right.  \label{a4}
\end{equation}
Here $\Psi $ is the scalar potential of the magnetic field; $\bf{d}$ is
a unit vector directed along the line of sight on to the observer; and $%
B_d(\theta ,\phi )$ is the distribution of the magnetic component specified
over the entire spherical surface. The distribution $B_d(\theta ,\phi )$
corresponds to two magnetograms of the visible and averted sides of the Sun
(it is assumed that there is a change of sign in the case of the back
magnetogram). Different ways of specifying boundary conditions on the
averted side will be discussed below. Two methods of specifying boundary
conditions on the outer surface (\ref{a4})-a) and (\ref{a4})-b) correspond
to the most frequently used typical settings. Condition (\ref{a4})-a)
corresponds to the magnetic field radiality, and condition (\cite{Altschuler}%
) (\ref{a4})-b) refers to a regular (at infinity) solution (such a setting
is frequently used in treatments of more sophisticated, combined magnetic
field models involving a potential-field model, either as a part of a
general model (\cite{Zhao and Hoeksema}), or as the initial approximation of
a general model (\cite{Aly and Seehafer} and \cite{Amary}).

We shall pursue the solution of this boundary-value problem in a standard
form of harmonic expansion in terms of eigensolutions of the Laplace
equation written in a spherical coordinate system. The spherical coordinate
system $(r,\theta ,\phi )$ is chosen here to conform to a Cartesian one, in
which the axis $z$ coincides with the direction $\bf{d}$. Such a choice,
as will be shown later in the text, simplifies greatly the solution of this
problem. For the sake of definiteness, the axis $x$ is assumed to lie in the
plane produced by the axis $z$ and the heliographic axis $z^{^{\prime }}$
(from here on we shall use primed symbols related to the heliographic
coordinate system) directed northward; the axis $y$ is taken to correspond
to a right-hand coordinate system.

\section{Method of solving the boundary-value problem}

As the starting point, we use the following form of harmonic expansion of
the scalar magnetic potential:

\begin{equation}
\Psi \left( r,\theta ,\phi \right) =R\sum\limits_{l=1}^{\infty \vee
L}\sum\limits_{m=-l}^lc_l^mf_l(r)\widetilde{P}_l^{|m|}\left( \cos \theta
\right) e^{im\phi },  \label{a5}
\end{equation}

(\ref{a5}) where $R$ is the solar radius, $c_l^m$ stands for the sought-for
complex coefficients of expansion

\begin{equation}
c_l^{-m}=\overline{c}_l^m,  \label{a6}
\end{equation}

$\widetilde{P}_l^m=P_l^m/\sqrt{2\pi w_l^m},P_l^m$ -- represents the Legendre
functions (\cite{Abramowitz and Stegun}), and

\begin{equation}
w_l^m=\int\limits_{-1}^1\left[ P_l^m\left( u\right) \right]
^2du=(l+1/2)^{-1}(l+m)!/(l-m)!  \label{a7}
\end{equation}
The expression (\ref{a5}) is an exact solution if the summation with respect
to the index $l$ is made ad infinitum, or, a finite approximation, if the
summating is limited to the value of $L$ (the main index of expansion). As
is customary, the contribution of the monopole term of expansion $l=0$ is
neglected in this case. Let the dependencies $f_l(r)$ be defined for two
cases of upper boundary conditions:

\begin{equation}
f_l(r)=\left\{
\begin{array}{c}
\frac 1{a_l-1}\left( a_l\overline{r}^{-l-1}-\overline{r}^l\right) \\
\overline{r}^{-l-1}
\end{array}
\right|
\begin{array}{c}
\ for\ case\ (4)-a) \\
for\ case\ (4)-b)
\end{array}
,  \label{a8}
\end{equation}
where $a_l=(R_s/R)^{2l+1}$, $\overline{r}=r/R$, and $R_s$ is the source
radius. In the form (\ref{a5}), each term of expansion satisfies the Laplace
equation (\ref{a2}) and the upper boundary conditions (\ref{a4}). We seek
the solution for (\ref{a5}) in a class of square-integrable functions on a
sphere $R$, which is equivalent to the boundedness condition of infinite
sums of the form:

\begin{equation}
\sum\limits_{l=1}^\infty \sum\limits_{m=-l}^{m=l}|c_l^m|^2<\infty  \label{a9}
\end{equation}
In accordance with the designations used here, we substitute (\ref{a5}) into
(\ref{a3}), multiply the resulting equality by $\widetilde{P}_l^{|m|}\left(
\cos \theta \right) e^{-im\phi }\cos \theta \sin \theta d\phi d\theta $ and
write the system of equalities in the form of equalities of integrals over
angular coordinates

\begin{eqnarray}
&&\ \ \ \ \ \int\limits_0^\pi \int\limits_0^{2\pi }B_d(\theta ,\varphi )%
\widetilde{P}_l^{|m|}\left( \cos \theta \right) e^{-im\phi }\cos \theta \sin
\theta d\phi d\theta  \label{a10} \\
\quad &=&\sum\limits_{l=1}^{\infty \vee L}c_k^m\int\limits_{-1}^1\left[
g_ku^2\widetilde{P}_k^{|m|}\left( u\right) \widetilde{P}_l^{|m|}\left(
u\right) +u(u^2-1)\frac{d\widetilde{P}_k^{|m|}}{du}\widetilde{P}%
_l^{|m|}\left( u\right) \right] du
\end{eqnarray}
Here $u=\cos \theta $ and $g_k=-\left( df/d\overline{r}\right) _{\overline{r}%
=1}>0$. Using the orthogonality property of Legendre polynomials and the
known recurrence relations for expressions of the form $uP_n^m$ and $%
(u^2-1)dP_n^m/du$ (\cite{Abramowitz and Stegun}), and upon introducing the
designation for the left-hand side of the equalities in (\ref{a10}) $b_l^m$
(these quantities will be called the weight coefficients), the expression (%
\ref{a10}) can be brought into the form
\begin{equation}
b_l^m=a_{-}^{m,l}c_{l-2}^m+a_0^{m,l}c_l^m+a_{+}^{m,l}c_{l+2}^m,  \label{a11}
\end{equation}
where
\begin{eqnarray}
a_{-}^{m,l} &=&\frac{w_{l-1}^m}{\sqrt{w_{l-2}^mw_l^m}}\frac{(l-m-1)(l+m)}{%
(2l-3)(2l+1)}(g_{l-2}+l-2)  \label{a12} \\
a_0^{m,l} &=&\left\{ \frac{w_{l+1}^m}{w_l^m}\left( \frac{l-m-1}{2l+1}\right)
^2(g_l+l)+\frac{w_{l-1}^m}{w_l^m}\left( \frac{l+m}{2l+1}\right)
^2(g_l-l-1)\right\}  \nonumber \\
a_{+}^{m,l} &=&\frac{w_{l+1}^m}{\sqrt{w_{l+2}^mw_l^m}}\frac{(l-m+1)(l+m+2)}{%
(2l+5)(2l+1)}(g_{l+2}-l-3)  \nonumber
\end{eqnarray}
In view of the property (\ref{a6}), for determining the coefficients $c_l^m$%
, it will suffice to exploit equation (\ref{a11}) for positive $m$ only. The
equalities (\ref{a11}) break down into $m$ independent systems of equations,
each of which is representable in a matrix form (for the finite expansion up
to $L$)
\begin{equation}
A_{ij}^mC_j^m=B_i^m,  \label{a13}
\end{equation}
where
\begin{eqnarray}
A_{ij}^m &=&\delta _{j+2}^ia_{-}^{m,2j+m-2}+\delta _j^ia_0^{m,2j+m-2}+\delta
_{j-2}^ia_{+}^{m,2j+m-2},  \label{a14} \\
C_j^m &=&c_{2j+m-2}^m,B_i^m=b_{2i+m-2}^m,1\leq i,j\leq (L-m)/2+1,\delta
_j^i=\left\{
\begin{array}{c}
1,i=j \\
0,i\neq j
\end{array}
\right.  \nonumber
\end{eqnarray}
For each value of $m$, the matrices $\widehat{A}^m$ in (\ref{a13}) are of
the tridiagonal form and are always nonsingular. The numerical solution of
matrix equations of such a form presents no special problems. Noteworthy are
two factors which reduced the search for the desired coefficients to a
simple scheme. This implies the selection of a special coordinate system and
the inclusion of the term $\cos (\theta )$ in the integrals of (\ref{a10}).
A similar procedure, which involves multiplying by an additional function $%
\sin (\theta ^{\prime })$, was implemented in a good-performance method of
reconstructing the magnetic field from the component $B_l$ (\cite{Hoeksema}%
), a component of the magnetic field along the line of sight at the time of
central meridian passage by the point of its determination. Searching the
harmonic expansion in \cite{Hoeksema} reduces to equations similar to (\ref
{a14}), with coefficients having the same qualitative character of behavior
depending on its indices. Although test calculations discussed below fully
justify the new constructive approach and make an impressive case in favor
of its correspondence to the boundary-value problem under consideration, we
wish to give a more rigorous mathematical rationale to the proposed method.

\section{Existence and uniqueness of the solution of the boundary-value
problem}

To prove the existence, it will suffice (assuming a square-integrability of
the distributions $\cos (\theta )B_d$) to show a uniform convergence in the
indices $L$ and $m$ of the inverse matrix norms $\left( \widehat{A}^m\right)
^{-1}$ involved in the expression (\ref{a13}). We confine ourselves to case (%
\ref{a4})-b) of upper boundary conditions where the structure of the
matrices $\widehat{A}^m$ becomes triangular (the elements above the
principal diagonal, corresponding to the coefficients $a_{+}$, are zero).
For triangular matrices, the elements of the principal diagonal are known to
be their eigenvalues. Since the norm of any nonsingular inverse matrix is $%
\left| \lambda _{\min }\right| ^{-1/2}$ ($\lambda _{\min }$ being the least
eigenvalue of the initial matrix), our problem implies analyzing minimum
elements of the principal diagonals of the matrices under consideration.
Diagonal elements for case (\ref{a4})-b) of upper boundary conditions have
the form:
\begin{equation}
a_0^{m,l}=\frac{(2k-1)^2}{2k+1}\frac{2m+1+2k}{2m+3+4k},\qquad
(l-m)=2k=0,1,2...  \label{a15}
\end{equation}
Straightforward examination of formula (\ref{a15}) reveals that the value of
$a_0^{m,l}$ independently on $m$ for any $L$ is minimum when $k=1$, i.e.
\begin{equation}
\lambda _{\min }^m=\frac 13\frac{2m+3}{2m+7}\geq \lambda _{\min }=\lambda
_{\min }^0=\frac 17  \label{a16}
\end{equation}
A uniform convergence of the desired norms both in the index $L$ and in the
index $m$ immediately follows from (\ref{a16}) (the norm of a full operator,
which transforms the entire infinite set of the coefficients $c$ with
positive indices $m$ to an infinite set of the coefficients $b$, is $\sqrt{7}
$). Thus the existence of the solution of the boundary-value problem with
upper boundary conditions of the form (\ref{a4})-b) is proved. In view of
the completeness of the functional space considered on a boundary sphere,
the uniqueness of the expansion (\ref{a5}) follows automatically (for $r=R$%
), with coefficients satisfying equations (\ref{a11}). The uniqueness of the
resulting solution throughout the region $\Omega $ , in turn, follows from
the uniqueness of a regular (at infinity) solution of the Dirichlet problem.

In the case of boundary conditions of the form (\ref{a4})-a) the elements $%
a_{+}$ from (\ref{a12}) are not equal to $0$, but with respect to elements
of the other diagonals, the character of behavior of which remains the same,
they are small as the index $l$ tends to infinity. Therefore, in this case,
too, the uniform convergence of each finite $m$ in $L$ seems to occur. It
has not yet been possible to find an explicit index expression in $m$ for a
minimum eigenvalue or for any lower edge of a set of eigenvalues in order to
prove a uniform convergence in this index. It will be assumed that there
exists at least one solution satisfying the boundary-value problem
formulated. In this case it is reasonably straightforward matter to prove
its uniqueness. Let us consider this proof.

Assume that there exists a second solution, and let the difference of two
solutions be designated by $\chi $. Then, for $\chi $, with the boundary
conditions (\ref{a4})-a) remaining the same, the boundary conditions (\ref
{a3}) become homogeneous. For uniqueness, it will suffice to show that the
solution of $\chi $ can be only trivial. Consider the inequality:
\begin{equation}
I=\int\limits_\Omega \left( \bf{\nabla }\left( \bf{d}\cdot \bf{%
\nabla }\chi \right) \right) \cdot \left( \bf{\nabla }\left( \bf{d}%
\cdot \bf{\nabla }\chi \right) \right) dV\geq 0.  \label{a17}
\end{equation}
For $I$, we write the following obvious chain of equalities:
\begin{eqnarray}
I &=&\int\limits_\Omega \bf{\nabla }\left[ \left( \bf{d}\cdot
\bf{\nabla }\chi \right) \cdot \bf{\nabla }\left( \bf{d}\cdot
\bf{\nabla }\chi \right) \right] dV-\int\limits_\Omega \left( \bf{d}%
\cdot \bf{\nabla }\chi \right) \cdot \Delta \left( \bf{d}\cdot
\bf{\nabla }\chi \right) dV  \label{a18} \\
&=&\int\limits_{S_p\cup S_s}\left( \bf{d}\cdot \bf{\nabla }\chi
\right) \left( \bf{n}\cdot \bf{\nabla }\left( \bf{d}\cdot
\bf{\nabla }\chi \right) \right) ds=\int\limits_{S_s}\left( \bf{d}%
\cdot \bf{\nabla }\chi \right) \frac \partial {\partial r}\bf{\nabla
}\left( \bf{d}\cdot \bf{\nabla }\chi \right) ds  \nonumber
\end{eqnarray}
Next, we substitute into (\ref{a18}) the expression for a unit vector $%
\bf{d}$ in terms of coordinate unit vectors of a spherical coordinate
system $\bf{d}=\bf{e}_z=\cos \theta \bf{e}_r-\sin \theta \bf{%
e}_\theta $, and extend the chain of equalities:
\begin{eqnarray}
I &=&\int\limits_{S_s}\left( \cos \theta \chi _r-\frac{\sin \theta }r\chi
_\theta \right) \left( \cos \theta \chi _{rr}-\frac{\sin \theta }r\chi
_{r\theta }+\frac{\sin \theta }{r^2}\chi _\theta \right) ds  \label{a19} \\
&=&\int\limits_{S_s}\cos \theta \chi _r\left( \cos \theta \chi _{rr}-\frac{%
\sin \theta }r\chi _{r\theta }\right) ds  \nonumber \\
&=&-R_s\int\limits_{-\pi }^\pi \int\limits_0^{2\pi }\cos \theta \chi
_r\left( 2\cos \theta \chi _r+\sin \theta \chi _{r\theta }\right) d\phi
d\theta
\end{eqnarray}
It is used in (\ref{a19}) that angular partial derivatives of any order of
the scalar potential are zero on the surface $S_s$ according to the equality
$\chi \left( R_s,\theta ,\phi \right) =0$. The dependence $\chi _{rr}$ is
expressed in terms of the first derivative $\chi _r$ from the Laplace
equation written in the spherical coordinate system. Furthermore, on
performing a change of variables $u=\cos \theta $ and introducing the
designation $Q(u)=\int\limits_0^{2\pi }\left( \chi _r\right) ^2d\phi $, we
continue transforming the integral:
\begin{eqnarray}
I &=&-R_s\int\limits_{-1}^1\left[ 2u^2Q-\frac 12\left( 1-u^2\right)
uQ_u\right] du  \label{a20} \\
&=&-R_s\int\limits_{-1}^1Q\left[ 2u^2+\frac 12\frac d{du}\left( 1-u^2\right)
u\right] du=-\frac{R_s}2\int\limits_{-1}^1Q\left( 2u^2+1\right) du  \nonumber
\end{eqnarray}
The expression under the integral sign in (\ref{a20}), in view of the
definition of $Q$, is always positive; whence it follows that $I\leq 0$. The
last inequality is compatible with the inequality (\ref{a17}) only when the
integral $I$ is zero. The following logical chain holds:

\begin{eqnarray*}
I &=&0\Rightarrow \left\{ \left[ \bf{d}\cdot \bf{\nabla }\chi \equiv
const\right] \wedge \left[ \bf{d}\cdot \bf{\nabla }\chi \equiv
0|_{S_p}\right] \right\} \Rightarrow \left\{ \bf{d}\cdot \bf{\nabla }%
\chi \equiv 0\right\} \\
&\Rightarrow &\left\{ \left[ \chi (\bf{r})=\chi (x,y)\right] \wedge
\left[ \chi =0|_{S_s}\right] \right\} \Rightarrow \chi \equiv 0
\end{eqnarray*}
Thus the uniqueness theorem is proved.

\section{The method of recurrence calculation of weight coefficients}

Preparatory to discussing the practical implementation of the proposed
method and results of tests calculations, we will first give some attention
to the relevant procedure of calculating weight coefficients. This procedure
provides essentially a high accuracy in numerical tests and in research work
with particular data. For the time being, we leave aside the problem of
populating with data the back side of the solar sphere. Imagine the
situation where we have two magnetograms of the visible and invisible sides
of the Sun, representing given values of the magnetic component at nodes of
a rectangular grid on the picture plane. In the initial stage we scale the
input data to a rectangular grid of the plane $(\theta ,\phi )$, having a
density which at least does not deteriorate the resolution of the input
data. Such a translation of the data can be performed, for example, on the
basis of a spline-representation of the input data in the picture plane.
Next, we determine the continuous distribution $B_d(\theta ,\phi )$ as an
approximation of the grid values by cubic local splines. The resulting
distribution is continuous up to the second partial derivatives with respect
to the indicated arguments. Let the full integrals of weight coefficients
defined by formula (\ref{a10}) be divided into elementary sums of integrals
over separate small rectangular regions corresponding to the size of the
initial grid subdivision. As the function $B_d(\theta ,\phi )$ is
represented by cubic polynomials inside of each elementary region, we shall
have to evaluate integrals of the form
\begin{equation}
\left( \int\limits_{\phi _i}^{\phi _{i+1}}\phi ^\alpha e^{-im\phi }d\phi
\right) \left( \int\limits_{\theta _j}^{\theta _{j+1}}\theta ^\beta
P_l^m(\cos \theta )\cos \theta \sin \theta d\theta \right) ,\qquad \alpha
,\beta =0,1,2,3.  \label{a21}
\end{equation}
The expression (\ref{a21}) involves two types of integrals. Integrals over $%
\phi $ have straightforward analytical expressions. Integrals over $\theta $
are also expressible analytically. With low indices $l$ and $m$ these
integrals can be calculated using the representation of Legendre polynomials
in power polynomials. With larger indices, however, such an approach leads
to a substantial accumulation of errors (as in the case of a calculation of
the polynomials themselves) and makes evaluations of the integrals
impossible. It is known that a most appropriate method of calculating
Legendre polynomials is to calculate them by recurrence formulas (see, for
example, \cite{Altschuler}). For the integrals appearing in (\ref{a21}), (by
a corresponding formal integration of recurrence equations for Legendre
polynomials) it is possible to obtain the respective recurrence relations
which make it possible to perform their calculations as effectively as when
calculating the polynomials. The above method of calculating the weight
coefficients (henceforth referred to as the recurrence analytical
integration) gives no Gibbs effect and, as a result, with minimum
computational inout, leads to solutions corresponding with a high accuracy
on the lower boundary to the initial spline-representation of the grid data,
as well as furnishing the opportunity to carry out, without perceptible
losses, multiple reconstructions of the magnetic field.

\section{Analysis of test calculations, and of calculations simulating the
handling of real magnetograms}

The pre-assigned a ''standard'' magnetic field in the form of the expansion (%
\ref{a5}) was used to analyze the performance of the method proposed above
when specifying, as a boundary condition, the data of real daily
magnetograms. To construct a ''standard'' expansion we used an arbitrary
magnetic synoptic map from Kitt Peak, specified on a grid with 1-degree
resolution. Next, in the heliographic coordinate system $\left( r^{\prime
},\theta ^{\prime },\phi ^{\prime }\right) $ using the method reported in
\cite{Hoeksema} for solving the boundary-value problem with a given
component $B_l$, we determined the coefficients of the expansion (\ref{a5})
up to $L=90$. To model the operation with real differently-resolved
magnetograms in the expansion of the standard magnetic field, the value of $%
L $ was decreased simply by discarding redundant expansion terms. In doing
this, proper account was taken of the correspondence of the typical scales
of the last expansion terms to the resolution being analyzed. Main attention
was given to three typical scales corresponding to the resolution of
magnetograms from Stanford ($\sim 15^{\circ }$), Mt.Wilson ($\sim 4^{\circ }$%
), and Kitt Peak ($\sim 1^{\circ }$, lowres). The standard field formed the
basis for constructing ''model daily magnetograms'' for a subsequent
extrapolation. Subsequent expansions were performed up to the same number of
terms. Final comparisons were made both with values of the full magnetic
vector and the standard field potential at different heights and with values
of the standard expansion coefficients of (\ref{a5}). For the latter, the
resulting field was re-expanded: the distribution of $B_r$ on the surface $%
S_p$ was calculated and reconstructed from a normal component in the
heliographic coordinate system. Precisely this transition from the
coordinate system $\left( r,\theta ,\phi \right) $ to the system $\left(
r^{\prime },\theta ^{\prime },\phi ^{\prime }\right) $ will be assumed
throughout in what follows where the comparison of harmonic coefficients is
implied. Note that in the final comparison there was no point in making
comparisons with the standard model pointwise, because each re-expansion of
the magnetic field employed the method of approximation at nodes by local
splines which yield inaccurate values at nodes of the initial grid, while
when different coordinate systems are used, the nodes themselves are
inaccurate. The key criterion involved comparing the configurations of
contour lines of the components under consideration and their location.
Results of subsequent extrapolations were considered ''equivalent'' to the
starting model if contour lines were impossible to distinguish visually from
those given by the standard magnetic field and if the difference in
positions of the contour lines did not exceed the initial resolution of the
fields that were reconstructed. Comparisons of harmonic coefficients were
made separately only up to $l=9$, while a complete survey involved comparing
plots of linear characteristics of the spectrum from the index $l$ (for each
$l$, quadratic sums of coefficients with different $m$ were calculated). In
this case, results were considered equivalent if spectral lines merged
together over the entire range of $l$. Particular attention was given to the
comparison of neutral lines on the source surface and to positions of their
intersection of the equator. As is known, neutral lines and intersection
points, a calculation of which is of great practical significance, are the
most sensitive to minor distortions of large-scale background magnetic
fields.

Test I. The standard magnetic field model was used to calculate (omitting
the phase of constructing magnetograms) uniformly in coordinates on a sphere
$\left( R,\theta ,\phi \right) $ the distribution of the $B_d$-component.
The same was done for the $B_l$- and $B_r$-components. Reconstructions were
performed from $B_d$, $B_l$ and $B_r$, respectively. All cases showed a
complete equivalence for all of the above-mentioned criteria.

Conclusion: Mathematically, all methods of specifying boundary conditions
for the reconstruction are equivalent in regard to the degree of accuracy --
quite a natural result.

Test II. Information contained in real magnetograms is given in the form of
values of the component $B_d$ which are uniformly distributed in the picture
plane. When nodal points are projected onto the surface of a sphere, as they
approach the limb, the resolution of information is deteriorated
substantially (the distance between adjacent nodes on the surface of the
sphere increases, which is equivalent to a loss of information). The tests
outlined below were carried out with the purpose of assessing the influence
of this property on the reconstruction of magnetic fields. Note that, when
extrapolating from $B_l$, the influence of this property is unimportant in
consequence of the use of synoptic magnetograms constructed on the basis of
sampling measurements only in the neighborhood of the central meridian.

The standard magnetic field was used to preliminarily calculate a direct and
back artificial magnetogram with the structure of the data which fully
corresponded to the structure of data recording in real magnetograms.
Subsequently, these magnetograms were treated as input material for
extrapolating the magnetic field. The results obtained showed that an
extrapolation on the basis of proposed method eliminates unfortunately the
possibility of using magnetograms corresponding to the resolution of the
Stanford magnetograms. Differences from a true model of the magnetic field,
associated with a nonuniform distribution of information across the solar
disk, in this case lead to an irreplaceable distortion of the resulting
model for all criteria. Such distortions are also distinguishable for
Mt.Wilson magnetograms, but they are significantly smaller, and, in
principle, corresponding real magnetograms can be used. An extrapolation
from artificial magnetograms corresponding to Kitt Peak magnetograms, showed
very good results. Nonuniformity effects of data resolution disappeared
almost totally in all their manifestations. The final model in this case is
equivalent to a true model, both in the comparison of contour lines and of
expansion coefficients. Note that they are in either case difficult to
distinguish from each other in comparisons of images of the $B_d$-components
of the initial standard magnetogram and of a newly reconstructed one.
However, when full images of any magnetic component is developed onto a
rectangle in the plane $\left( \theta ^{\prime },\phi ^{\prime }\right) $,
if nonuniformity effects of data resolution are essential, one can see a
clear boundary separating the direct and back magnetograms, as well as a
distortion zone around it. For the Kitt Peak case such a boundary was
virtually undetectable, and on the source surface all contour lines merged
totally together.

Conclusion: There exist limitations caused by the resolution of the
magnetograms used. The proposed method produces faithful results starting
with the resolution of Kitt Peak magnetograms and higher.

\section{The problem of specifying a back magnetogram, and results of
calculations from real magnetograms}

First we present some qualitative considerations regarding the consequences
caused by the scarcity of data corresponding to the back magnetogram. On the
photospheric surface the influence of the back side on the main strong
small-scale field can have a substantial effect only in the neighborhood of
the limb corresponding to the field scale. With an increase in the altitude,
weaker components of a larger-scale field starts to play a crucial role.
Therefore, the region (henceforth referred to as the confidence region),
which is the least subjected to the influence of the back side, may be
represented as a region bounded by a dome-shaped surface with the height $R$%
. If the purpose of research is limited only to examinations inside such a
region, then it is in principle not very important how the back magnetogram
is filled. It is possible, for example, to assume zero boundary conditions
or to symmetrically represent the values of the magnetogram being analyzed.

If it is proposed to most fully represent the real boundary conditions of
the back side in order to minimize the distortions near the confidence
region boundary or to ascend significantly above this region to the source
surface, then the first natural possibility would be to endeavor to
replenish information for the back side using a corresponding synoptic
magnetogram (additionally there exists the hopeful way of using of the
vector magnetograms but it connects to principally another method -- the
regularization method \cite{Gary}, the discussion of this possibility is
beyond the scope of this work). For instance, it is possible to reconstruct
a full model of the magnetic field from the $B_l$-component, and to
calculate a back magnetogram. In a sense, such a procedure is justified. In
this case main attention is directed to the large-scale field which is
relatively little affected by substantial changes during a solar rotation.
Therefore, one would expect a good correspondence between large-scale fields
on the source surface obtained as a result of extrapolations from synoptic
data and magnetogram data. Also, it would be desirable to derive new
information from the new structural elements associated with the current
state of the field. In case of favorable results, it would be possible for
us to be able, in particular, to predict changes in solar wind
characteristics. Already first calculations of this kind revealed two
characteristic types of distortions. One is the limb distortion (a
pronounced small but intense band of distortions at the interface of the
magnetograms). It is apparent that this type is associated with the
impossibility of reconciling (at the interface) of the direct and back
magnetograms constructed as indicated above. The other type is the
distortion ''plateau''. This type is distinguished only in the analysis of a
rectangular image of the structure of the $B_r$-component on the source
surface. It is particularly well seen in the three-dimensional image of the
reconstructed distribution of $B_r\left( R_s,\theta ^{\prime },\phi ^{\prime
}\right) $. In this case it is clearly seen that two regions corresponding
to the direct and back magnetograms are detached from each other and
uniformly displaced from each other along the functional axis $B_r$.
Furthermore, if this gap is artificially eliminated, then the two regions
corresponding to the positions of the direct and back magnetograms would
gradually go over into each other, and the structure of $B_r\left(
R_s,\theta ^{\prime },\phi ^{\prime }\right) $ would always agree well with
that obtained by reconstructing the magnetic field from a full synoptic map
(from the component $B_l$). The gap does typically not exceed 10\% of a
maximum value of $B_r$ and can differ in sign. The plateau effect is unseen
on the lower surface because it in its magnitude corresponds to very weak
values of the field. Conceivably this effect represents an inaccuracy of
specifying the zero value of the source magnetograms. This effect is
extremely complicated to eliminate. If we try to eliminate the gap merely by
the addition of a small constant value to the data of the direct or back
magnetograms, then one is left with the possibility that the data from both
magnetograms are shifted by the same amount. This involves a relatively
complicated transformation of the monopole expansion term which we discard.
In such procedures, it varies and can transform to other expansion terms
which are taken into account. Each such fitting entails a new reconstruction
of fields, which demands much computational time. Numerical calculations
showed that it is not always possible to achieve a correspondence between
results of reconstructions from synoptic data and full-disk data. In the
case of such global structures of $B_r\left( R_s,\theta ^{\prime },\phi
^{\prime }\right) $, the arrangement of neutral lines can differ
substantially. The positions of important equator intersecting points of
neutral lines are most strongly responsive to such modifications.
Intersection points are almost always coincident in comparative calculations
of $B_l$-based reconstructions from Stanford and Kitt Peak magnetograms.
Since their calculations are based on totally different measurements in
regard to both resolution and time, these points may be thought of as being
stable in a sense and reflecting the physical state of large-scale fields
which is eventually associated with current conditions of the interplanetary
field at the Earth's orbit. It is therefore very important to achieve a
relative stability of these points in calculations from daily magnetograms.

A next step involved an attempt to determine the back magnetogram using,
instead of a synoptic magnetogram, some set of daily magnetograms covering
the whole surface of the Sun over a time interval incorporating the
magnetogram under consideration. The principle of construction is as
follows. To determine the $B_d$-component at an arbitrary point on the solar
surface, a set of magnetograms containing this point was used. It was
assumed that the axis of solar rotation is normal to the plane of the
Earth's orbit. Within such an approximation, for a particular point it is
possible to determine the desired $B_d$-component based on two components
from two arbitrary magnetograms for the same point. Finally, $B_d$ at the
desired point was determined by an average over the values calculated from
all possible pairs of magnetograms. Thus a full synoptic map of the $B_d$%
-component was generated. Interestingly, when a full $B_d$-synoptic
magnetogram was generated from a set of magnetograms appearing in the file
heading of a usual $B_l$-synoptic map, all calculations for these two
magnetograms by respective methods yielded virtually identical results from
comparisons on both the lower and upper surfaces. In particular, the most
sensitive neutral lines merged entirely together. This showed that the two
types of synoptic magnetograms are virtually identical to each other as
regards their information content. Naturally, the new maps contained neither
limb- nor plateau-distortions. It is clear that the above-mentioned
uncertainties of zero counts of the data from daily magnetograms are
averaged when generating synoptic maps of either type and give no
perceptible effects. When a part of the $B_d$-synoptic magnetogram was
substituted for by the data from the daily magnetogram, limb distortions
were vanishingly small. This was rather surprising and showed that the
effort made in this area was not useless and that the possibilities for a
further improvement of the situation have not been exhausted. Unfortunately,
although the plateau-distortions decreased slightly, they were still large
enough to be able to affect the arrangement of intersection points. The zero
uncertainty of a direct magnetogram is not fully compensated by an averaged
back magnetogram. Neutral lines of the source can also sometimes be affected
markedly when the number of daily magnetograms used to construct a $B_d$%
-synoptic magnetogram is varied. Thus the solution of the problem treated
here rests on the insufficiently good quality of daily magnetograms
employed. Kit Peak daily magnetograms made available for general use are
essentially meant for research into strong fields; moreover, they are not
the initial product of measurements. Initial measurements are with
significantly higher resolution. It is hoped that not all possibilities of
improving the determination of the zero count have been exhausted, both in
regard to initial the measurements and at the stage of their conversion to
final usable data. Unfortunately, this author has no technical way of
handling raw measurements in order to continue analysis along this line.

To illustrate the performance of the new method we give one example of a
calculation of open magnetic field regions from a daily magnetogram. By the
open magnetic field region is meant an area of the solar surface, with all
outgoing field lines reaching the source surface. To identify such region
involves calculating all field lines starting from points which are
uniformly distributed on the source surface. Crossings of field lines with
the photosphere give images of such regions. Figs. 1 and 2 present two
calculations from a daily magnetogram (Fig. 1) and from a corresponding
synoptic $B_l$-magnetogram (Fig. 2). Footpoints of open field lines here are
shown by circles superposed on soft X-ray images of the Sun. The light tone
corresponds to coronal hole regions usually associated with regions of open
field lines. Calculations from the daily magnetogram are an obvious
advantage. The situation illustrated here is typical of the other cases
considered in this study. Calculations from daily magnetograms, as a rule,
adequately depict the structure of most visible coronal holes, which is a
great improvement over synoptic calculations. An important point is that
calculations depend only slightly on the method of specifying the back
magnetogram. The dependence on ''back'' boundary conditions has a
substantial influence only on the position of the upper ends of field lines
issuing out of open regions; furthermore, the field line configuration
starts to change noticeably only at altitudes larger than the solar radius.

\section{Conclusions}

Main results and conclusions may be summarized as follows.

-- A new method has been developed for extrapolating the solar magnetic
fields from daily full-disk magnetograms.

-- A mathematical rationale for the new method was given.

- An analysis was made of the applicability of the method, with due regard
for the particular characteristics of existing data.

- The most suitable technique has been found for specifying the boundary
conditions on the back side of the Sun.

- It has been shown that this method can be used successfully only when
high-resolution daily magnetograms are employed (with the exception of
Stanford magnetograms).

- Calculations confirmed that in the confidence region defined above, one
can expect a good correspondence between calculated and real magnetic field
structures.

- In contrast to other techniques, this method makes it possible to describe
current conditions of a global magnetic structure (not averaged over time
and not local).

- The analysis made in this paper has revealed hidden limitations of
high-resolution magnetograms which manifest themselves when reconstructing
the magnetic structure on the source surface. It may be concluded that with
the proviso that they are eliminated or their influence is adequately
reduced, it is possible to predict a change of sign of the near-terrestrial
interplanetary magnetic field from the reconstructed picture of the radial
magnetic field on the source surface.

\begin{acknowledgements}

I am grateful to Dr ~J. ~Harvey for valuable discussions.
I thank to Mr.~V.~G.~Mikhalkovsky for his assistance in preparing the
English version of this report.

This work was supported within the State Scientific Research program on
''Astronomy''as well as grant 00-02-16456 of the Russian Foundation 
for Fundamental Research.

NSO/Kitt Peak generate in collaboration with NSF/NOAO, NASA/GSFC \& NOAA/SEL
data used here .
\end{acknowledgements}

\newpage

\begin{center}
{\Large {Figure captions} }
\end{center}

Figure 1. Calculation of regions of open field lines from the daily
magnetogram of November 27, 1993, $L=30$, SXT-27/11/93

Figure 2. Calculations of regions of open field lines from synoptic
magnetogram CR-1876, $L=30$, SXT-27/11/93

\end{article}


\begin{thebibliography}{Wang, Y.-M. and N.R. Sheeley, 1992}
\bibitem[Altschuler et al., 1977]{Altschuler}  Altschuler, M.D., Levine,
R.H., Stix, M., and Harvey, J.: 1977, \textit{Solar Phys.}, \textbf{51}, 345.

\bibitem[Hoeksema, 1984]{Hoeksema}  Hoeksema, J.T.: 1984, \textit{Structure
and Evolution of The Large Scale Solar and Heliospheric Magnetic Fields}.,
\textit{Ph. D. Diss}., Stahford Univ..

\bibitem[Wang, Y.-M. and N.R. Sheeley, 1992]{Wang and Sheeley}  Wang, Y.-M.
and N.R. Sheeley, Jv.:1992, \textit{Astrophys. J.}, \textbf{392}, 310.

\bibitem[Zhao and Hoeksema, 1994]{Zhao and Hoeksema}  Zhao, X.P. and
Hoeksema, J.T.:1994 \textit{Solar Phys.}, \textbf{151}, 91.

\bibitem[Aly and Seehafer, 1993]{Aly and Seehafer}  Aly, J.J. and Seehafer,
N.: 1993, \textit{Solar Phys.,} \textbf{144}, 243.

\bibitem[Amary et al, 1997]{Amary}  Amary, T., Aly, J.J., Luciany, J.F.,
Boulmezaoud, T.Z. and Mikic, Z.: 1997 \textit{Solar Phys.,} \textbf{174},
129.

\bibitem[Abramowitz and Stegun 1964]{Abramowitz and Stegun}  Abramowitz, M.
and Stegun, I.A.: 1964, \textit{Handbook of Mathematical Functions. Natl., }%
Bureau of Standards.

\bibitem[Gary, 1996]{Gary}  Gary, G.A.:1996 \textit{Solar Phys.,} \textbf{163%
}, 44.
\end{thebibliography}
\end{document}